\documentstyle[12pt,psfig]{article} 
\begin{document}
\begin{flushright}
%CUPP-00/5 \\
\texttt{hep-ph/0107022} 
\end{flushright}
\vskip 5pt
\begin{center}
{\Large {\bf Violation of the Equivalence Principle in the light
of the SNO and SK solar neutrino results}}
\vskip 20pt

\renewcommand{\thefootnote}{\fnsymbol{footnote}}

{\sf Amitava Raychaudhuri 
\footnote{E-mail address: amitava@cubmb.ernet.in}} 
and 
{\sf Arunansu Sil 
\footnote{E-mail address: arun@cubmb.ernet.in}}  
\vskip 10pt  
{\it Department of Physics, University of Calcutta, 92
Acharya Prafulla Chandra Road, \\ Kolkata 700009, India.}\\ 
\vskip 15pt
{\bf Abstract}
\end{center}

{\small 
The SNO result on charged current deuteron disintegration, the
SuperKamiokande 1258-day data on electron scattering, and other
solar neutrino results are used to revisit the model of neutrino
oscillations driven by a violation of the equivalence principle.
We use a $\chi^2$ minimization technique to examine oscillation
between the $\nu_e$ and another active neutrino, both massless,
and find that within the Standard Solar Model the fit to the SNO
and SuperKamiokande spectra are moderately good while a very good
fit is obtained when the absolute normalizations of the $^8$B and
$hep$ neutrino fluxes are allowed to vary. The best fit prefers
large, but not maximal, mixing, essentially no $hep$ neutrinos,
and a 40\% reduction in the $^8$B neutrino flux. The fit to the
total rates from the different experiments is not encouraging but
when the rates and spectra are considerd together the situation
is much improved.  We remark on the expectations of the VEP model
for the neutral current measurements at SNO. } 
\vskip 20pt 
\begin{center}
PACS NO. 26.65.+t, 14.60.Lm, 14.60.Pq, 04.80.Cc
\end{center}

\vskip 20pt

%\newpage
%\renewcommand{\thesection}{\Roman{section}}
\renewcommand{\thefootnote}{\arabic{footnote}}
\setcounter{footnote}{0}

%%%%%%%%%%%%%%%%%%%%%%%%%%%%%%%%%
\section{Introduction}
%%%%%%%%%%%%%%%%%%%%%%%%%%%%%%%%%

The Sudbury Neutrino Observatory (SNO) collaboration has
presented results from 241 days running of their solar neutrino
experiment \cite{sno1}, a reminder that it is still early days
yet in this fruitful area of astro-particle physics. From the
data on charged current (CC) deuteron disintegration ($\nu_e + d
\rightarrow  p + p + e^-$) they observe a $\nu_e$ flux which is
$3.3\sigma$ less than the neutrino flux measured by the
Super-Kamiokande collaboration (SK) {\em via} $\nu - e^-$
scattering (ES) \cite{sk1258}\footnote {SNO also has results from
ES events \cite{sno1} which agree with the SK results
\cite{sk1258} but is of poorer statistics currently.}.  This
further strengthens the oscillation interpretation of the solar
neutrino problem and, at least in a two neutrino scenario,
disfavours the mixing of the $\nu_e$ with a sterile species.

Violation of the Equivalence Principle (VEP) has been examined in
the literature  as a possible solution to the solar neutrino
problem \cite{mrs, kuo, gago, casini, bahkra}, one of the
alternatives to the mass-mixing vacuum oscillation and the matter
induced Mikheyev-Smirnov-Wolfenstein (MSW) approaches \cite{bf}.
In this note we revisit the VEP solution in a two-flavor picture
with the $\nu_e$ oscillating to another active species both of
which are assumed to be massless.

Neutrino oscillation in VEP can occur if (a) the weak equivalence
principle is not satisfied, i.e.,  the coupling of neutrinos to
the gravitational field is nonuniversal and (b) the flavor
eigenstates are not identical to the states that couple to
gravity \cite{gasp}. {\em It does not require neutrinos to carry
a non-zero mass.} The weak equivalence principle demands the
coupling of particles to the ambient gravitational potential
$\phi$ to be of the form $f \phi E$, where $E$ is the particle
energy, and $f$ a universal coupling constant. If the latter
varies from one neutrino species to another then that would
constitute a violation of the equivalence principle. If $f_1 \neq
f_2$ in a two-neutrino framework, then these states define a
basis in the two-dimensional space which, in general, could be
different from the flavor basis. 
\begin{equation}
\nu_e = \nu_1 \cos \theta + \nu_2 \sin \theta; \;\;\;\;
\nu_x = - \nu_1 \sin \theta + \nu_2 \cos \theta,
\label{eq:mix}
\end{equation}
where $x = \mu$, $\tau$.
The effect due to a small splitting $\Delta f$ will manifest
itself in the form of flavor oscillations, the wavelength going
to infinity as $\Delta f$ tends to zero.  The essential difference
between this approach and the mass-mixing solution appears in the
energy dependence of the survival probability.  For a
two-neutrino picture, the general expression for the survival
probability for an initial $\nu_e$ after propagation through a
distance $L$ is given by:
\begin{equation}
P_{ee}(E_\nu,L) = 1 - \sin^22\theta \sin^2
\left( \frac {\pi L}{\lambda}\right),
\label{eq:sprob}
\end{equation}
where, for the VEP induced case, the oscillation length
$\lambda$  is:
\begin{equation}
\lambda = \frac {2 \pi}{E_\nu \phi \Delta f}.
\label{eq:veplam}
\end{equation}
$E_\nu$ is the neutrino energy. Thus, unlike the mass-mixing case
where $\lambda \propto E_\nu$, for VEP $\lambda \propto 1/E_\nu$.
Due to the different energy dependences of the survival
probability in the mass-mixing and the VEP alternatives, their
predictions can be quite different for solar neutrinos of
different energies. We follow the prevalent practice of choosing
the gravitational potential, $\phi$, to be a constant over the
neutrino path. This is the case if the potential due to the Great
Attractor \cite{dress} dominates over that due to the sun and
other heavenly bodies in our neighborhood. In such an event,
writing $\Delta F = \phi \Delta f/2$, the expression for the
oscillation wavelength, Eq. (\ref{eq:veplam}), becomes
\begin{equation}
\lambda = \frac{6.20 \times 10^{-13}
{\rm m}}{\Delta F}
\left(\frac{1 {\rm MeV}}{E_\nu}\right).
\label{eq:lammag}
\end{equation}

An alternate physics scenario which also leads to neutrino
oscillations with $\lambda$ $\propto 1/E_\nu$ is a picture of
violation of special relativity (VSR) \cite{coleman}.  If special
relativity is violated, the maximum attainable speed of a
particle {\em in vacuo} need not universally be the speed of
light $c$. In particular, if the maximum possible velocities of
the two types of neutrinos be $v_1$ and $v_2$ and these {\em velocity
eigenstate} neutrinos be related to the $\nu_e$ and $\nu_x$
through a mixing angle $\theta$ (see Eq.(\ref{eq:mix})) then in
this case the expression for $\lambda$ is:
\begin{equation}
\lambda = \frac {2 \pi}{E_\nu \Delta v}, 
\label{eq:vsrlam}
\end{equation}
where $\Delta v$ is the velocity difference for the neutrinos
$\nu_1$ and $\nu_2$. Comparing Eqs. (\ref{eq:vsrlam}) and
(\ref{eq:veplam}) one finds that the energy dependence of the
oscillation length is identical in the two cases\footnote{It has
been shown that inclusion of CPT-violating interactions in
addition to Violation of Special Relativity can lead to more
general energy dependences involving $1/E_\nu$, $E_\nu$, and
constant terms \cite{CPT}}  and the role of $\Delta v$ in the VSR
case is the same as that of $\phi \Delta f$ in the VEP formalism.
Here, we use the terminology of the VEP mechanism but the results
can be taken over {\em mutatis mutandis} to the VSR situation.

In this work we use the BBP2000 calculation \cite{bp2000} of the
solar neutrino flux as the SSM reference. In addition to this, we
explore the possibility of the absolute normalizations of the
solar $^8$B- and $hep$-neutrino spectra\footnote{The SNO and SK
experiments are sensitive only to the $^8$B and $hep$ neutrinos.}
being different from their BBP2000 SSM predictions.  If $X_B$ and
$X_{hep}$ denote the factors by which the absolute normalizations
are multiplied, we use the data to find the best-fit values for
these. We find that all the fits are improved in a noteworthy
manner when $X_{B}$ is permitted to be less than unity.

%%%%%%%%%%%%%%%%%%%%%%%%%%%%%%%%%
\section{Update of VEP for the SK ES results}
%%%%%%%%%%%%%%%%%%%%%%%%%%%%%%%%%

In contrast to other recent work on VEP \cite{gago, casini} based
on the 825-day SK spectral data, in a previous publication
\cite{mrs}, the SK ES results from 1117 days of running
\cite{sk1117} were considered. It was found that the predictions
were by-and-large robust. One trend, which was remarked upon, was
that the softening of the highest energy SK ES data had resulted
in a reduction of $X_{hep}$.  In fact, it was noted that the
best-fit to the day-night spectrum of the 1117-day SK data
preferred a negligibly small value of $X_{hep}$.

For a further update, we have performed a similar analysis with
the SK 1258-day data. The SK results are presented in the form of
number of events (normalized to the SSM expectation) in 19
electron recoil energy bins of width 0.5 MeV in the range 5 MeV
to 14 MeV and a 19th bin which covers the events in the range 14
to 20 MeV \cite{sk1258}. 

The definition of $\chi^2$, the error correlations, {\em etc.}
are chosen in the same manner as in \cite{mrs}. Suffice it to say
that we have included the statistical error, the uncorrelated
systematic errors, and the energy-bin-correlated experimental
errors \cite{skspec} as well as those from the calculation of the
shape of the expected spectrum \cite{shape}.  When we allow the
normalizations of the $^{8}$B and $hep$ fluxes to vary, we do not
include their astrophysical uncertainties separately. The results
are presented in Table 1.

\begin{center}
\begin{tabular}{|c|c|c|c|c|c|c|c|}
\hline
Fitted&Case &$\sin^2 2\theta$ & $\Delta F$ & $X_B$ & $X_{hep}$ &
$\chi^2_{min}$/d.o.f. & g.o.f \\
Data&&& $(10^{-24})$ & & & & (\%)\\
\hline
SK1258 ES&1a&0.72 & 0.22 & 1.0& 1.0 & 29.10/17&3.37 \\
Day-Night averaged&&&&(fixed) & (fixed) && \\ \cline{2-8}
Spectrum&1b&1.0 & 0.99 & 0.77 & $2.4\times10^{-5}$&11.38/15&72.52 \\
&&&&& && \\ \hline
SK1258 ES&1c&0.72 & 0.21 & 1.0& 1.0 & 43.71/36&17.67 \\
separate Day \& Night&&&&(fixed) & (fixed) && \\ \cline{2-8}
Spectra&1d&1.0 & 0.98 & 0.77 & $9.3\times10^{-8}$&24.68/34&87.92\\
&&&&& && \\ \hline
\end{tabular}
\end{center}

\begin{description}
\item{\small \sf Table 1:} {\small \sf The best-fit values of the
parameters, $\sin^22\theta$, $\Delta F$, $X_B$, $X_{hep}$, the
$\chi^2_{min}$, and the goodness of fit (g.o.f.) for fits to the
SK1258 ES spectra.}
\end{description}

We find that the fits are by-and-large robust; using the separate
day and night data or the day-night averaged data affects them
only marginally. Further, the best-fit parameters have not
changed much and they are well within the 90\% C.L. allowed
regions of the previous analysis using the 1117-day data.  For
example, for the fit to the the day-night averaged data sample,
in  the SSM case ($X_B = X_{hep} = 1$), the best-fit values of
$\sin^22\theta$ and $\Delta F$ have changed $<$ (5 - 10)\% but
the quality of fit has significantly dropped (g.o.f. = 76.7\% for
the fit to the 1117-day data). When $X_B$ and $X_{hep}$ are
allowed to vary, the best-fit value of (a) $X_{hep}$ is tiny, and
(b) $\Delta F$ is larger than earlier and closer to the value for
the SSM fit, a welcome feature. The goodness of fit is also a bit
lower.  For the fit to the separate day and night data, the
differences between the fits to the 1258-day  and 
1117-day data broadly follow the same pattern as above.

%%%%%%%%%%%%%%%%%%%%%%%%%%%%%%%%%
\section{VEP Fits to the SNO CC and SK ES spectra}
%%%%%%%%%%%%%%%%%%%%%%%%%%%%%%%%%

The SNO collaboration has published the first measurement of the
spectrum of electron kinetic energy produced from CC deuteron
disintegration by solar neutrinos. The results are presented in
eleven electron kinetic energy bins ranging from 6.5 MeV to 13.0
MeV. This provides a new check on solutions to the solar neutrino
puzzle. The best-fit values of the VEP parameters obtained from
fitting this data sample are shown in Table 2.

When the SNO CC and SK 1258-day ES data (separate day and night
or day-night averaged) are simultaneously addressed within the
VEP picture,  the best-fit values are remarkably close to each
other in the SSM fits (2a), (2c), and (2e) though the goodness of
fit is low.  When $X_B$ and $X_{hep}$ are allowed to vary, then
the fits to the combined SK ES and SNO CC spectra improve. These
results are also presented in Table 2.

\begin{center}
\begin{tabular}{|c|c|c|c|c|c|c|c|}
\hline
Fitted&Case&$\sin^2 2\theta$ & $\Delta F$ & $X_B$ & $X_{hep}$ &
$\chi^2_{min}$/d.o.f. & g.o.f \\
Data&&& $(10^{-24})$ & & & & (\%)\\
\hline
&2a&0.70 & 0.20 & 1.0& 1.0 & 9.74/9&37.19 \\
SNO CC&&&&(fixed) & (fixed) && \\ \cline{2-8}
Spectrum&2b&0.72 & 0.20 & 1.09 & $6.6\times10^{-7}$&9.70/7&20.62 \\
alone&&&&& && \\ \hline
&2c&0.72 & 0.21 & 1.0& 1.0 & 42.04/28&4.30 \\
SNO CC \& SK1258 &&&&(fixed) & (fixed) && \\ \cline{2-8}
(Day-Night averaged)&2d&0.66 & 0.99 & 0.62 &
$1.9\times10^{-10}$&24.60/26&54.17 \\
Spectra&&&&& && \\ \hline
&2e&0.72 & 0.21 & 1.0& 1.0 & 56.60/47&15.93 \\
SNO CC \& SK1258 &&&&(fixed) & (fixed) && \\ \cline{2-8}
(Separate Day \& Night)&2f&0.66 & 0.99 & 0.61 &
$9.3\times10^{-10}$&37.79/45&76.84\\
Spectra&&&&& && \\ \hline
\end{tabular}
\end{center}

\begin{description}
\item{\small \sf Table 2:} {\small \sf The best-fit values of the
parameters, $\sin^22\theta$, $\Delta F$, $X_B$, $X_{hep}$, the
$\chi^2_{min}$, and the g.o.f. for fits to SNO CC and SK1258 ES
spectra.}
\end{description}

As seen in the fit to the SK ES spectra alone (Table 1), here
again the fits do not change whether the separate day and night
data or the day-night averaged data are used. In view of this, in
the next section, we use only the day-night averaged data.

We show in Fig. 1, the allowed regions for the parameters
$\sin^22\theta$ and $\Delta F$ at 90\% C.L. obtained by fitting
the SK 1258-day ES spectrum (dot-dashed line), the SNO CC
spectrum (small-dashed line), and for a combined fit to the SNO
CC and the SK ES spectra (solid -- for the SSM case -- and dotted
-- when $X_B$ and $X_{hep}$ are allowed to vary). It is
noteworthy that for the SSM, the allowed regions from the
different fits overlap very nicely.

%%%%%%%%%%%%%%%%%%%%%%%%%%%%%%%%%
\section{VEP Fits to total rates and the SNO CC and SK ES spectra}
%%%%%%%%%%%%%%%%%%%%%%%%%%%%%%%%%

The SK ES and SNO CC spectrum measurements are excellent filters
for choosing between alternative solutions of the solar neutrino
problem. But it has to be borne in mind that they are both
sensitive only to the highest energy $^8$B neutrinos. Further
constraints on any putative solution come from the total solar
neutrino rates measured by the Chlorine and Gallium experiments.
In order to be acceptable, the VEP solution must also be
confronted with these results, to which we now turn.  The data
used in the $\chi^2$ analysis of total rates are given in Table
3.

\newpage
\begin{center}
\begin{tabular}{|c|c|c|c|c|}
\hline
Experiment & Chlorine & Gallium & SK ES & SNO CC \\ \hline
$\frac{\rm Observed \;\;
Rate}{\rm SSM \;\; Prediction}$ & $0.337 \pm 0.030$ & $0.584 \pm
0.039$ & $0.4594 \pm 0.016$ & $0.347 \pm 0.029$ \\ \hline
\end{tabular}
\end{center}
 
\begin{description}
\item{\small \sf Table 3:} {\small\sf The ratio of the observed
solar neutrino rates to the corresponding BBP2000 SSM predictions
used in this analysis. The results are from Refs. \cite
{solar,sk1258,sno1}.The Gallium rate is the weighted average of
the Gallex, SAGE, and GNO results.}
\end{description}

The best-fit parameters obtained by fitting the total rates are
presented in Table 4 (sets 4a and 4b). Here the goodness of fits
are not high. In order to ascertain how the total rate
measurements mesh with the SK ES and SNO CC measurements, we have
performed combined fits to the rates and spectral data. These
results are also shown in Table 4. Note that the SSM is
disfavoured by this analysis.

\begin{center}
\begin{tabular}{|c|c|c|c|c|c|c|c|}
\hline
Fitted&Case&$\sin^2 2\theta$ & $\Delta F$ & $X_B$ & $X_{hep}$ &
$\chi^2_{min}$/d.o.f. & g.o.f \\
Data&&& $(10^{-24})$ & & & & (\%)\\
\hline
&4a&1.0 & 1.66 & 1.0& 1.0 & 5.50/2&6.39 \\
Total Rates&&&&(fixed) & (fixed) && \\ \cline{2-8}
alone&4b&1.0 & 1.66 & 0.74& 1.0 & 2.40/1&12.13 \\
 &&&&  & (fixed) && \\ \hline
SNO CC, SK1258 &4c&0.72 & 0.21 & 1.0& 1.0 & 117.62/32&$1.0\times
10^{-9}$ \\
ES spectra \& Total &&&&(fixed) & (fixed) && \\ \cline{2-8}
Rates (incl.&4d&1.0 & 1.60 & 0.77 &
$6.8\times10^{-8}$&33.54/30&29.96 \\
SNO \& SK)&&&&& && \\ \hline
SNO CC, SK1258 &4e&0.72 & 0.21 & 1.0& 1.0 & 92.43/30&$2.8\times
10^{-6}$ \\
ES spectra \& Total &&&&(fixed) & (fixed) && \\ \cline{2-8}
Rates (excl.&4f&1.0 & 1.62 & 0.77 &
$1.8\times10^{-5}$&31.04/28&31.53 \\
SNO \& SK)&&&&& && \\ \hline
\end{tabular}
\end{center}

\begin{description}
\item{\small \sf Table 4:} {\small \sf The best-fit values of the
parameters, $\sin^22\theta$, $\Delta F$, $X_B$, $X_{hep}$, the
$\chi^2_{min}$, and the g.o.f. for fits to the total rates along
with the SNO CC and SK ES spectra.}
\end{description}

When we fit the total rates and the SNO CC and SK ES spectra
together, the SNO and SK total rates and the corresponding
spectra cannot be considered to be statistically independent
pieces of data. However, their exclusion from the fits would also
not be entirely correct as the fitted spectral ratios
(observed/SSM) do not completely determine the total rates. We
have therefore chosen to examine two extreme alternatives; i.e.,
in the fits to the spectra and the total rates together, (a) in
one case we include the SK and SNO total rates in the fit, and
(b) in the other case we have excluded them. It is gratifying
that the best-fit values of the parameters of the two cases (4c,
4e) or (4d, 4f) are not very different, a reflection of the
comparative energy independence of the (observed/SSM) spectra.

The best-fit values of the combined total rates + spectra fits
for the SSM case (4c, 4e) are within the 90\% C.L. allowed
regions obtained from fits to the spectra alone (see Fig. 1).

%%%%%%%%%%%%%%%%%%%%%%%%%%%%%%%%%
\section{SNO NC expectation, Comparison of different fits}
%%%%%%%%%%%%%%%%%%%%%%%%%%%%%%%%%

SNO will soon start taking data for a calorimetric measurement of
neutral current (NC) deuteron disintegration ($\nu + d
\rightarrow  \nu + p + n$) to which all active neutrinos
contribute equally. A measure of neutrino oscillations is
provided by the ratio of the NC and CC rates, $R_{NC}$ and
$R_{CC}$, which is somewhat less sensitive to theoretical
uncertainties than the rates themselves. We define
\begin{equation}
R_{SNO} = \frac{R_{NC}}{R_{CC}}.
\end{equation}
For the best fit values of parameters obtained in the earlier
sections, we present the predicted values of $R_{SNO}$ in Table
5. For an assessment of the various fits, for every set of
best-fit parameters we also show in Table 5 the following
quantities: the calculated rates for the Chlorine, Gallium, SK ES
and SNO CC observations, and also the $\chi^2$ obtained for the
calculated SK ES and SNO CC electron spectra.

\begin{center}
\begin{tabular}{|c|c|c|c|c|c|c|c|}
\hline
Case& Cl & Ga
&\multicolumn{2}{|c|}{SK}&\multicolumn{3}{|c|}{SNO}\\ \cline{4-8} 
& & & Rate  & $\chi^2$ & Rate &$\chi^2$ &$R_{SNO}$ \\ 
& ($0.337 \pm 0.030$) & ($0.584 \pm 0.039$) & ($0.4594 \pm
0.016$) & (spect) & ($0.347 \pm 0.029$)& (spect) & \\ \hline\hline
1a&0.516 & 0.940 & 0.479&29.10 &0.353&11.65 & 1.00\\ \hline
1b&0.437 & 0.796 & 0.462&11.38 &0.380&20.96 & 0.71\\ \hline
1c&0.508 & 0.940 & 0.476&29.40 &0.341&10.73 & 1.04\\ \hline
1d&0.436 & 0.798 & 0.458&11.50 &0.380&21.62 & 0.71\\ \hline\hline
2a&0.522 & 0.941 & 0.502&45.50 &0.356&9.74 & 0.99\\ \hline
2b&0.523 & 0.942 & 0.522&78.60 &0.356&9.70 & 1.07\\ \hline
2c&0.510 & 0.940 & 0.481&29.96 &0.343&10.35 & 1.03\\ \hline
2d&0.496 & 0.858 & 0.461&13.14 &0.418&12.53& 0.51\\ \hline
2e&0.506 & 0.940 & 0.479&30.25 &0.336&10.15 & 1.05\\ \hline
2f&0.495 & 0.859 & 0.459&13.20 &0.416&12.56& 0.51\\ \hline\hline
4a&0.443 & 0.609 & 0.605&355.6 &0.520&23.18 & 0.67\\ \hline
4b&0.339 & 0.596 & 0.448&23.60 &0.385&12.37 & 0.67\\ \hline
4c&0.511 & 0.940 & 0.482&30.26 &0.344&10.29 & 1.03\\ \hline
4d&0.352 & 0.613 & 0.456&18.40 &0.400&12.94& 0.67\\ \hline
4e&0.511 & 0.940 & 0.480&29.75 &0.344&10.44 & 1.03\\ \hline
4f&0.350 & 0.608 & 0.456&18.40 &0.400&12.97& 0.67\\ \hline
\end{tabular}
\end{center}

\begin{description}
\item{\small \sf Table 5:} {\small \sf The total rates for the
different experiments obtained by using the best-fit values of
the VEP parameters presented in Tables 1, 2, and 4.  Also
presented are the $\chi^2$ for fits to the observed SK ES and SNO
CC spectra using these values of the parameters. The predictions
for $R_{SNO}$ are also shown.}
\end{description}

From Table 5 it can be seen that fits to the spectra by themselves
give best-fit parameters grossly disfavoured by the total rates
measured in the Chlorine and Gallium experiments. The most acceptable
results are obtained in cases (4d, 4f) -- combined fits to the SK ES
spectra, SNO CC spectra, and the Chlorine and Gallium total rates
measurement with $X_B$ = 0.77 (see Table 4).  These values of
parameters will be further strengthened if $R_{SNO}$ is measured
around 0.67.

%%%%%%%%%%%%%%%%%%%%%%%%%%%%%%%%%
\section{Conclusions and Discussions}
%%%%%%%%%%%%%%%%%%%%%%%%%%%%%%%%%

In this work, we have considered the VEP oscillation explanation
of the solar neutrino problem within a two-flavor (active)
scenario in the light of the SNO CC measurement, the SK 1258-day
ES data, and the total rates from the Chlorine and Gallium
experiments.

We find that the VEP fits to the SK ES data are robust. The
best-fit parameters  have changed only marginally from the
earlier fit to the 1117-day data.

In the literature there are many variants of the data-fitting
procedure for solar neutrinos. The solution preferred by Nature
should fit all the solar neutrino results collectively {\em and
also individually}. A good fit to one (or some) of the results
may do rather poorly when confronted with another piece of datum.
A good fit to all the data taken together may mask bad fits to
some of the results taken in isolation. This caveat is overlooked
in much of the current literature. We have attempted to fit the
SK ES and SNO CC spectra individually and jointly, and also the
total rates by themselves and jointly with the spectral data in
an attempt to find a region of parameter space that is acceptable
in every respect. These results are given in Table 5 and in
Tables 1, 2, and 4. We find that the best-fit parameters of the
cases (4d) and (4f) (see Tables 4 \& 5) best meet the requirements
alluded to above. Note that these fits correspond to $X_B \sim
0.77$ and predict $R_{SNO} \sim$ 0.67.

During the passage of the neutrinos from their point of
production to the solar surface, interactions with the ambient
matter, responsible for the MSW effect, become important. In the
presence of VEP and this MSW contribution, the effective neutrino
mass matrix in flavor space takes the form
\begin{equation}
M =\frac{1}{2} \left| \begin{array}{cc}
E_\nu\Delta F \cos 2\theta  - 2\sqrt{2} G_F n_e(r) &  E_\nu\Delta F
\sin 2\theta  \\
E_\nu\Delta F \sin 2\theta  & - E_\nu\Delta F \cos 2\theta 
\end{array} \right|,
\label{eq:mmat}
\end{equation}
where we have dropped an irrelevant part proportional to the
identity matrix. The MSW contribution in (\ref{eq:mmat}) inside
the sun turns out to be several orders of magnitude larger than
the terms due to VEP that we have discussed in this work. Recall
that we have assumed the neutrinos to be massless. For maximal
mixing (e.g., the preferred 4d and 4f cases), there is no
resonance effect and, in fact, till such time that the neutrino
emerges from the sun, the MSW contribution controls the  masses
in (\ref{eq:mmat}). Inside the sun, the $\nu_e$ is, therefore,  a
mass eigenstate to a very good approximation. The effect of VEP
oscillations begins to manifest itself only from then onwards.

The VEP oscillation wavelengths favored by the data are
comparable to the distance of the sun to the earth. Therefore,
seasonal effects are expected in this picture. When more 
seasonal data from SK accumulates, it will provide a good check
on this scenario.

Since the publication of the first results by the SNO
collaboration, a number of analyses of the data in the
mass-mixing vacuum oscillation and MSW approaches have appeared
\cite{bf}. The quality of these fits are comparable to those
found in the VEP picture. Here too, the parameters which best fit
the rates alone are somewhat different from those that best fit
the rates together with the spectra \cite{bf}. Further sharpening
of the data from new  and ongoing experiments, it is hoped,
will help to distinguish between these alternatives.

\vskip 30pt

{\large{\bf {Acknowledgements}}}\\

The authors are grateful to the Calcutta University Computer
Centre for the use of their Origin 2000 computer. They thank
Debasish Majumdar for his help.  The research of A.R. is
supported in part by a grant from CSIR, India while A.S.
enjoys a fellowship from UGC, India.

\vskip 30pt
%\newpage

\vskip 30pt
\newpage
\begin{figure}[thb]
%\begin{center}
%\vskip -2.00in
%\hskip -1.15in
%\psfig{figure=fig1.ps,xsize=0.5,ysize=0.5} 
%\vskip -4.10in
\psfig{figure=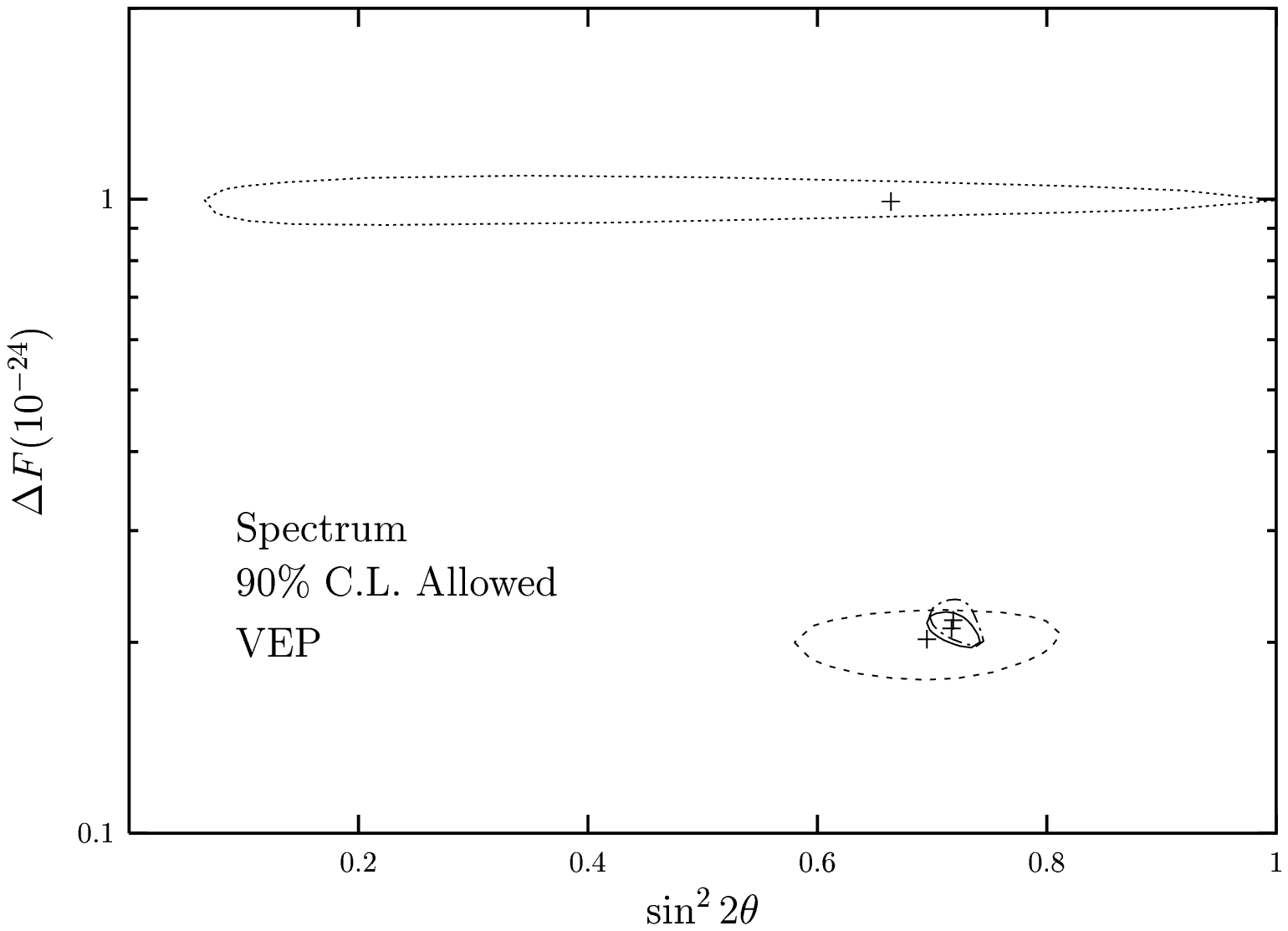,width=14.0cm,height=22.0cm} 
\vskip -3.50in
%\hskip -2.00in
\caption{\sf \small   
The 90\% C.L. allowed regions in the $\sin^2 2\theta$ - $\Delta
F$ plane. The area enclosed by the solid (dotted) line is allowed
by the SNO CC + SK1258 ES spectral data for the SSM (the model
where $X_B$ and $X_{hep}$ are allowed to vary).  The area allowed
by the SNO CC (SK1258 ES) spectrum alone for the SSM is enclosed
by the small dashed (dot-dashed) line. The best fit points have
been indicated.} %\end{center}
\end{figure}

\end{document}